\documentclass{iaus}
\usepackage{graphicx}
\usepackage{graphics}
\usepackage{amsmath}

\title[Planetary Transits and Tidal Evolution] 
{Planetary Transits and Tidal Evolution}

\author[Brian Jackson, Rory Barnes, \& Richard Greenberg]   
{Brian Jackson$^1$, Rory Barnes$^1$,
 \and Richard Greenberg$^1$}

\affiliation{$^1$Lunar and Planetary Laboratory, University of Arizona \\
1629 E University Blvd, Tucson AZ 85721-0092 USA 
\\ email: {\tt bjackson@lpl.arizona.edu}}

\pubyear{2008}
\volume{253}  
\pagerange{119--126}
\jname{Transiting Planets}
\editors{}
\begin{document}

\maketitle

\begin{abstract}
Transiting planets are generally close enough to their host stars that tides may govern their orbital and thermal evolution of these planets. We present calculations of the tidal evolution of recently discovered transiting planets and discuss their implications. The tidal heating that accompanies this orbital evolution can be so great that it controls the planet's physical properties and may explain the large radii observed in several cases, including, for example, TrES-4. Also because a planet's transit probability depends on its orbit, it evolves due to tides. Current values depend sensitively on the physical properties of the star and planet, as well as on the system's age. As a result, tidal effects may introduce observational biases in transit surveys, which may already be evident in current observations. Transiting planets tend to be younger than non-transiting planets, an indication that tidal evolution may have destroyed many close-in planets. Also the distribution of the masses of transiting planets may constrain the orbital inclinations of non-transiting planets.

\\ \\ Submitted 2008 Jul 3
\end{abstract}

\firstsection 
\section{Introduction}

\indent Most close-in planets, and thus most transiting planets, have likely undergone significant evolution of their orbits since the planets formed and the gas disks dissipated. Jackson et al. (2008a) showed that the initial eccentricities of close-in planets were likely distributed in value similarly to the eccentricities of planets far from their host stars. Current eccentricities, as well as reduced semi-major axes, result from subsequent tidal evolution. Here we apply similar calculations of tidal evolution to transiting planets discovered more recently. In all cases, orbital evolution has been significant.\\
\indent Tides heat planets as they change their orbits. Jackson et al. (2008b) computed the heating that accompanies tidal circularization, showing that many close-in planets experience large and time-varying internal heating. They noted that tidal heating has been large enough recently enough that it may explain the anomalously large radii of many close-in planets. Since then, as more transiting planets have been discovered, the number of planets with anomalously large radii has increased. Here we compute the tidal-heating histories for recently discovered transiting planets. \\
\indent Tidal evolution can also affect the probability for transits to be observable from Earth because it changes the orbits. As a result, tidal evolution may introduce biases into transit observations as we discuss in Section 4 below. 

\section{Orbital Evolution}
\begin{figure}[h]
\begin{center}
  \includegraphics[width=3.4in]{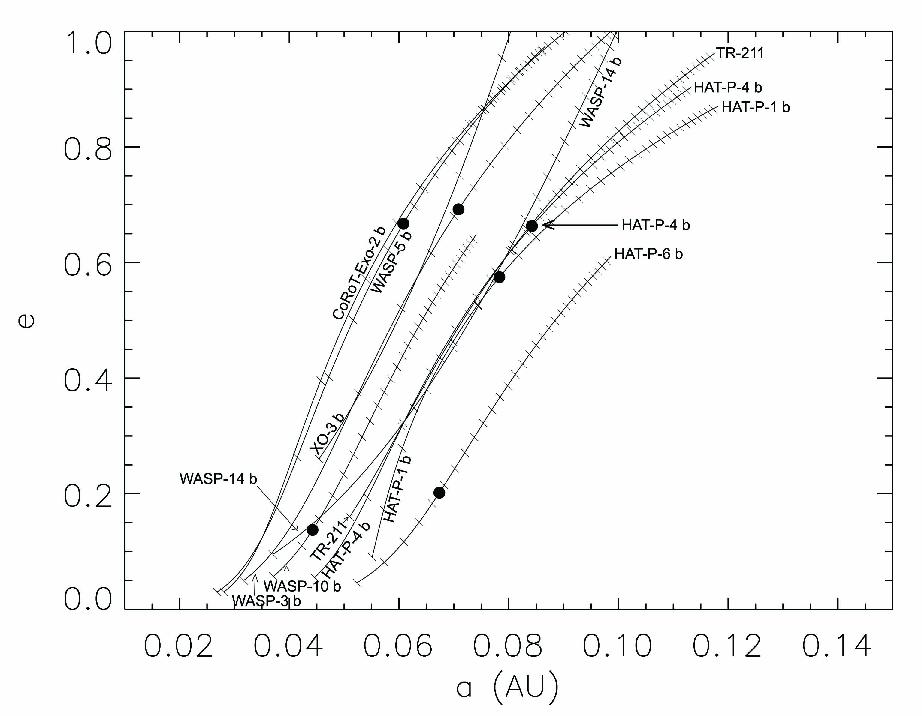} 
 \caption{Orbital evolution of all transiting planets with best-fit $e \neq 0$, discovered since \cite{Jacksonetal08a}. Current orbital elements are at the lower left end of each path. Tick marks are spaced at intervals of 500 Myr. Black dots show the age of each system for which an estimate is available. The age of OGLE-TR-211 b (called TR-211 in the figure) is not known; the black spot near the corresponding trajectory is for HAT-P-4 b. Orbital and physical parameters are taken from \cite{Alonsoetal08}, \cite{Bakosetal07a}, \cite{Kovacsetal07}, \cite{Noyesetal08}, \cite{Udalskietal08}, \cite{Pollaccoetal08}, \cite{Andersonetal08}, \cite{Christianetal08}, \cite{Joshietal08}, and \cite{Winnetal08}. With $a = 0.1589$ AU, HD 17156 b (\cite{Gillonetal08}) is off the right-hand side of the plot and experiences negligible tidal evolution.}
   \label{Fig1}
\end{center}
\end{figure}

\indent The distribution of eccentricities $e$ of extra-solar planets is very uniform over the range of semimajor axis values $a > 0.2$ AU (Jackson et al. 2008a). Values for $e$ are relatively large, averaging 0.3 and broadly distributed up to near 1. For $a < 0.2$ AU, eccentricities are much smaller (most $e < 0.2$), a characteristic widely attributed to damping by tides after the planets formed and the protoplanetary gas disk dissipated. However, estimates of the tidal damping often consider the tides raised on the planets, while ignoring the tides raised on the stars. Results depend on assumed specific values for the planets' poorly constrained tidal dissipation parameter $Q_p$. Perhaps most important, the strong coupling of the evolution of $e$ and $a$ is often ignored. \\
\indent \cite{Jacksonetal08a} integrated the coupled tidal evolution equations for $e$ and $a$ over the estimated age of many close-in planets. There we found that the distribution of initial (i.e. immediately after completion of formation and gas-disk migration) $e$ values of close-in planets do match that of the general population if stellar and planetary $Q$ values are $10^{5.5}$ and $10^{6.5}$, respectively, however the results are nearly as good for a wide range of values of the stellar $Q$. The accompanying evolution of $a$ values shows most close-in planets had significantly larger $a$ at the start of tidal evolution. The earlier gas-disk migration did not bring most planets to their current orbits. Rather, the current small values of $a$ were only reached gradually due to tides over the lifetime of the systems.\\
\indent Here we update those calculations by including more recently discovered transiting planets. The results highlight the importance of accounting for both the effects of the tide raised on the star and for the strong coupling between the evolution of semi-major axis and eccentricity. The results also show that neglecting either of these effects can lead to incorrect inferences about transiting (or other) close-in planets, such as beliefs that orbits must have been circularized. \\
\indent Fig. \ref{Fig1} shows the past tidal evolution of recently discovered transiting planets. We include only planets for which a best-fit non-zero eccentricity has been published. We show the tidal evolution of $e$ and $a$, calculated as described by \cite{Jacksonetal08a}, going back in time from their current nominal orbital elements. \\

\indent As noted by \cite{Jacksonetal08a}, the concavity of the $e$-$a$ trajectories reflects the dominant tide, either the tide raised on the planet, or the tide raised on the star. Where a trajectory is concave down (usually when $e$ is large), the effects of the tide raised on the planet dominates. Where a trajectory is concave up (usually when $e$ is small), the tide raised on the star dominates. In the latter case orbital angular momentum is transferred to the host star's rotation, resulting in a spin-up of the star (such as may be the case for $\tau$ Boo b (\cite{Henryetal00}). In most of the trajectories in Fig. \ref{Fig1}, there is a transition from dominance by the tide on the planet to dominance by the tide on the star, as $e$ passes from large ($> 0.4$) to small values. In these cases, neglecting the effects of the tide raised on the star would underestimate the rate of tidal evolution of $e$ and $a$, especially later in the evolution, when eccentricities are small. \\
\indent In fact, ignoring the coupling of $a$ and $e$ evolution has been an implicit feature of a very common approach to estimating the damping of $e$ values. There are numerous examples in the literature, as reviewed by \cite{Jacksonetal08a}, of computing and applying a "circularization timescale", $e/(de/dt)$, which is incorrectly assumed to apply over a significant part of the tidal evolution. The actual change of $e$ over time is often quite different from these "circularization timescale" considerations, due to the coupled changes in $a$, as discussed by \cite{Jacksonetal08a}. \\
\indent The problematic "circularization timescale" approach continues to be inappropriately applied for constraining the orbits of recently discovered transiting planets (e.g., \cite{Collier-Cameronetal07}, \cite{ODonovanetal07}, \cite{Gillonetal07}, \cite{Pontetal07}, \cite{Bargeetal08}, \cite{Burkeetal08}, \cite{Christianetal08}, \cite{Johnsonetal08}, \cite{JohnsKrulletal07}, \cite{McCulloughetal08}, \cite{Nutzmanetal08}, \cite{Weldrakeetal08}, and \cite{Joshietal08}). Consider the example of HAT-P-1 b. \cite{Johnsonetal08} obtained an orbital fit to the observations that admits an eccentricity as large as 0.067. However, on the basis of an estimate of the "circularization timescale", they then assumed that $e = 0$. Their estimate for the circularization timescale, based on an assumed $Q_p = 10^6$, is 0.23 Gyr, which is less than 10\% of the estimated age (2.7 Gyr) of the system. Based on those numbers, it is reasonable to expect the current $e$ to be several orders of magnitude smaller than its initial value. Even if we choose the larger value of $Q_p = 10^{6.5}$ suggested by the results of \cite{Jacksonetal08a}, the circularization timescale only increases to 0.6 Gyr, still short enough that $e$ should have damped to less than 0.01 during the lifetime of this system. (If we also include the effect of tides raised on the star, the circularization timescale is even shorter, although this factor is negligible in this particular case.) But recall that the circularization-timescale approach does not take into account the coupled tidal evolution of $a$ along with $e$. In effect, the decrease in a over time means that $e/(de/dt)$ cannot be treated as a constant. According to Fig. \ref{Fig1}, when the coupling of $a$ and $e$ is taken into account, $e$ could have started at $<$ 0.6 (an unremarkable value for a typical extra-solar planet), and it would still have a value of 0.09 (the best-fit value previously reported by \cite{Bakosetal07a}) at the present time. Clearly the circularization-timescale method, as it is commonly used, can drastically over-estimate the damping of orbital eccentricities. For HAT-P-1 b and many other cases, it has incorrectly constrained eccentricity values.\\
\indent At best, the circularization timescale $e/(de/dt)$ can only describe the current rate of evolution, so it is not relevant to the full history of the tidal evolution. Even if one is only interested in the current rate, there are pitfalls. One, of course, is uncertainty in the appropriate values of $Q$. For example, in a discussion of WASP-10b, \cite{Christianetal08} found the persistence of substantial eccentricity to be surprising. Christian et al. calculated a circularization timescale substantially less than 1 Gyr for $Q_p = 10^5$ to $10^6$. In fact, with $Q_p = 10^6$, the damping timescale is about 1 Gyr, and with $Q_p = 10^{6.5}$, it is 3 Gyr, so it is not clear why the observed value (about 0.057) is surprising. The age of this system also has been estimated at ~1 Gyr, so the current eccentricity would only be potentially problematic if $Q_p$ had the unlikely value of $10^5$.\\
\indent Another pitfall involves neglecting the effect of tides raised on the star by the planet, as was done by Johnson et al. and by Christian et al., and indeed in most of the papers cited above. For HAT-P-1 b they are not important. But for WASP-10 b, using the favored values $Q_p = 10^{6.5}$ and $Q_* = 10^{5.5}$, we find that including the effect of tides on the star, the circularization timescale drops from 3 Gyr to 0.5 Gyr. Evidently, the damping may be dominated by tides raised on the star. This dominance is also evident in Fig. \ref{Fig1}, where the evolution trajectory has been concave upward since the formation of the system. Also, in this case the timescale for damping $e$ is similar whether we simply use the circularization-timescale estimate or account for the coupled evolution of $e$ and $a$ (Fig. \ref{Fig1}), because the system happens to be young. More generally, however, it is a mistake to rely on circularization-timescale estimates. Rather it is essential to include the full coupled equations for tidal evolution.\\
\indent Based on estimates of short "circularization timescales", observers commonly assume $e = 0$ in fitting the orbits of close-in planets, rather than reporting observational upper limits on $e$. In fact, $e$-damping is probably much slower, as shown in Fig. \ref{Fig1}. Thus those observers fitting data to possible orbits should not discount substantial $e$-values on the basis of uncertain tidal models, but should solve for the best fit (and range of uncertainty) to their observations. Those results will help constrain the histories of these systems. Moreover, even small $e$-values may have important implications for the physical properties of the planets as discussed in the next section.\\
\indent In the above discussion we focus on two papers that provide examples of how misleading it may be to neglect tides raised on the star or the coupling of the evolution of $a$ with $e$. We cite these particular cases because the circumstances happen to provide good illustrative examples. However, it should be understood that those authors were following what has come to be a widespread standard procedure as observers try to reduce the range of uncertainty of orbital elements implied by their data. All observers should adopt a more complete tidal model before over-interpreting possible constraints.

\section{Tidal Heating}
\indent Extra-solar planets close to their host stars have likely undergone significant tidal evolution since the time of their formation. Tides probably dominated their orbital evolution once the dust and gas cleared away, and as the orbits evolved, there was substantial tidal heating within the planets. The tidal heating history of each planet may have contributed significantly to the thermal budget governing the planet's physical properties, including its radius, which in many cases may be measured by observing transits. Typically, tidal heating first increases as a planet moves inward toward its star and then decreases as its orbit circularizes. \\
\indent \cite{Jacksonetal08b} computed the plausible heating histories for several planets with measured radii, using the same tidal parameters for the star and planet that have been shown to reconcile the eccentricity distribution of close-in planets with other extra-solar planets. For several planets whose radii are anomalously large, we showed how tidal heating may be responsible. For one case, GJ 876 d (\cite{Riveraetal05}), tidal heating may have been so great as to preclude its being a solid, rocky planet. We concluded that theoretical models of the physical properties of any close-in planet should consider the possible role of tidal heating, which is time-varying, and can be quite large.\\
\indent Here we present heating rates and histories for transiting planets discovered more recently, and then discuss their implications. First, we calculate the possible tidal heating of planets whose eccentricities are reported to be zero (often because of the problematic circularization timescale discussed in Section 2 above). Because observations allow the possibility of non-circular orbits, we calculate tidal heating histories assuming various small but non-zero current eccentricities. Even very small current eccentricities ($<$ 0.01) may have dramatic implications in many cases. Second, we calculate heating histories of planets for which there is some non-zero eccentricity reported. We show that tidal heating is likely large in many cases and should be incorporated into physical models of planetary radii. We show that tidal heating may help reconcile discrepancies between predicted and observed planetary radii in many cases.

\subsection{Planets with reported zero eccentricities}
\begin{figure}[t]
\begin{center}
\includegraphics[width=3.4in]{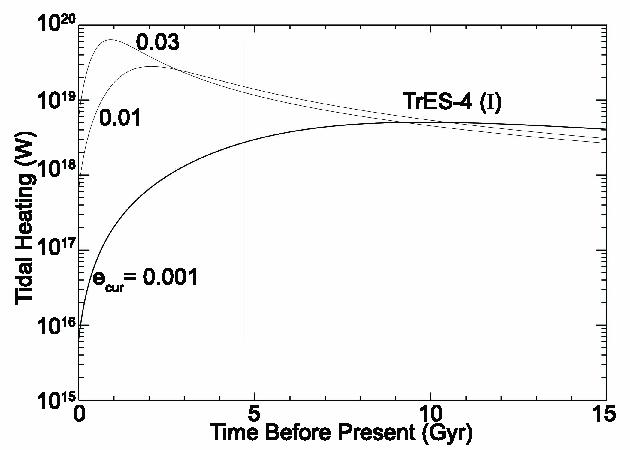} 
\caption{Tidal heating for TrES-4. Orbital and physical parameters are taken from \cite{Mandushevetal07}.}
\label{Fig2}
\end{center}
\end{figure}

\indent Fig. \ref{Fig2} shows the tidal heating history of TrES-4, the transiting planet with the largest radius identified to date \cite{Mandushevetal07}. We have computed these rates as described by \cite{Jacksonetal08b}. Although Mandushev et al. assume $e = 0$ for this planet, we show here (as for all the planets discussed in this section) heating curves for which we assume current $e$-values of 0.001, 0.01, and 0.03. The current heating rate (left edge of the graph) is greatest for the largest $e$ and smallest for smallest $e$, so the curves are readily identified. If it is available, an estimate for the age of the planet is shown as a vertical line. The ``(I)" next to the planet's name indicates that the observed radius is reported to be inflated relative to a theoretical prediction that neglected tidal heating. For planets that are reported not to be inflated, in subsequent plots, we put an ``(N)" next to the name. 

\begin{figure}[b]
\begin{center}
\includegraphics[width=3.4in]{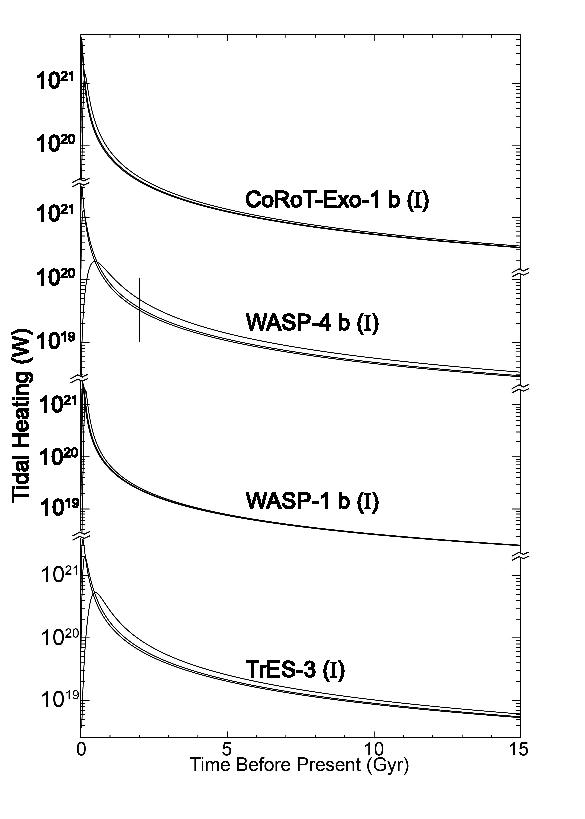} 
 \caption{Tidal heating for CoRoT-Exo-1 b, WASP-4 b, WASP-1 b, and TrES-3, all of which are reported to be inflated. Orbital and physical parameters are taken from \cite{Bargeetal08}, \cite{Wilsonetal08}, \cite{Collier-Cameronetal07}, and \cite{ODonovanetal07}, respectively.}
\label{Fig3}
\end{center}
\end{figure}
\begin{figure}[b]
\begin{center}
 \includegraphics[width=3.4in]{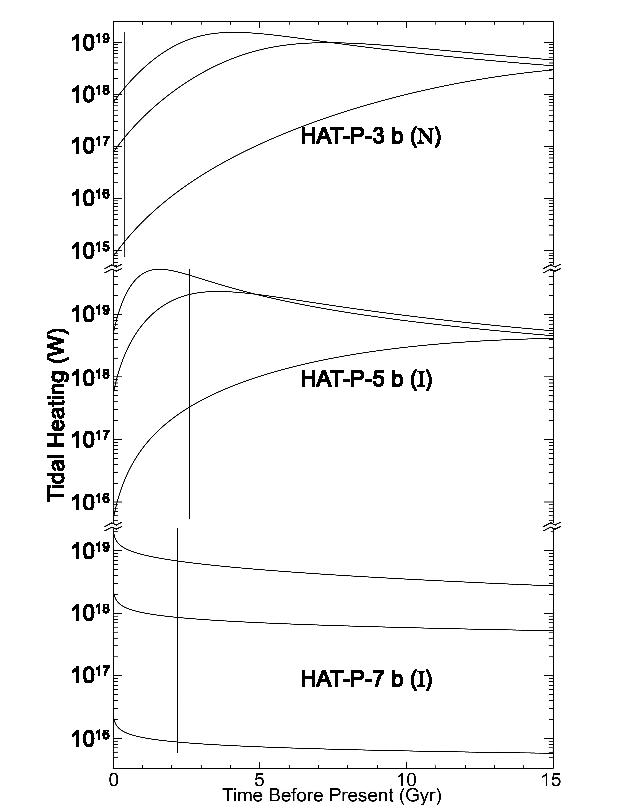} 
 \caption{Tidal heating of HAT-P-3 b, -5 b and -7 b. Orbital and physical parameters are taken from \cite{Torresetal07}, \cite{Bakosetal07c}, and \cite{Paletal08}, respectively.}
\label{Fig4}
\end{center}
\end{figure}
\begin{figure}[b]
\begin{center}
\includegraphics[width=3.4in]{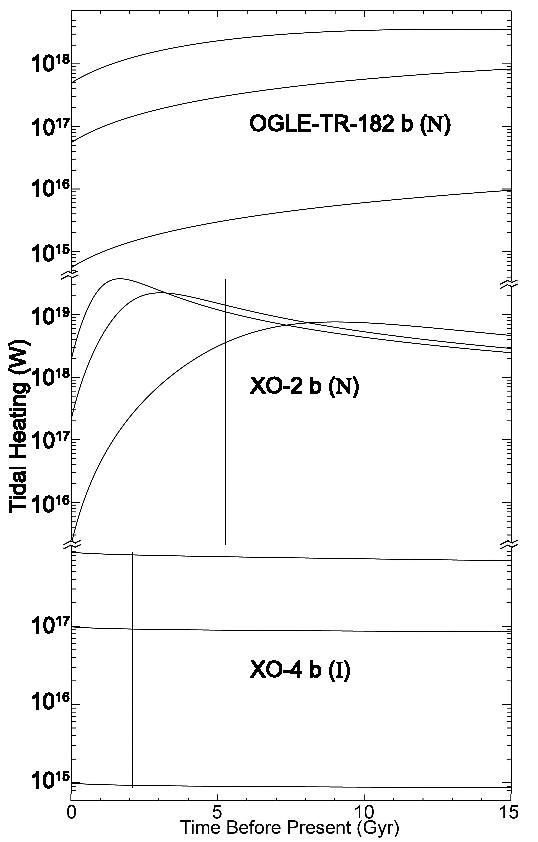} 
\caption{Tidal heating of OGLE-TR-182 b, XO-2 b and XO-4 b. Orbital and physical parameters are taken from \cite{Pontetal07}, \cite{Burkeetal08}, and \cite{McCulloughetal08}, respectively.}
\label{Fig5}
\end{center}
\end{figure}

\indent As shown in Fig. \ref{Fig2}, for TrES-4 even if the current $e < 0.01$, tidal heating could have been $> 10^{19}$ W for much of the past billion years. That much recent heating may be sufficient to pump up the radii of close-in extra-solar planets, as discussed by \cite{Jacksonetal08b}. Thus, it may help explain the anomalously large radius of TrES-4. \cite{Liuetal08}, motivated by \cite{Jacksonetal08b}, obtained similar results, although they assumed a current $e$ of 0.04 and $Q_p = 10^5$. \\
\indent The shape of TrES-4's heating curve is characteristic of many planetary heating curves. As tidal dissipation of orbital energy reduces $a$, the heating rate increases (time goes forward towards the left in these graphs). However, as $a$ drops, the rate of orbital circularization also increases. Eventually $e$ drops enough that tidal heating slows. (See Equation 1 from \cite{Jacksonetal08b}.) This increase and subsequent decrease in the heating rate results in a peak in the heating for many planets, including TrES-4. \cite{Jacksonetal08c} showed that the peak occurs at $e = 0.34$ for cases where tides on the planet dominate the tidal evolution and orbital angular momentum is conserved.\\
\indent Figs. \ref{Fig3} through \ref{Fig5} show (similarly to Fig. \ref{Fig2}) the heating for several other transiting planets whose reported $e = 0$. In the cases of CoRoT-Exo-1 b, WASP-4 b, WASP-1 b, TrES-3, HAT-P-5 b, and HAT-P-7 b (Figs. \ref{Fig3} and \ref{Fig4}), the radii have been reported to be inflated. Our results show recent (within the last Gyr) tidal heating exceeds $10^{19}$ W even for the small current eccentricities we consider. In these cases, it seems that tidal heating may contribute significantly to the inflated radii.

\indent For HAT-P-3 b and OGLE-TR-182 b, the radii are reportedly not inflated. The tidal heating rates in Figs. \ref{Fig4} and \ref{Fig5} do not allow the tidal heating to exceed $10^{19}$ W at any time during the past billion years, which may be consistent with the uninflated radius. \\
\indent HAT-P-3 b's radius is so small, in fact, that its discoverers (\cite{Torresetal07}) suggest that it may have a rocky core with a mass about 75 Earth masses. In the absence of tidal heating, such a large, rocky core would probably account for the small radius. However, such a large core could reduce the planet's effective $Q_p$ by orders of magnitude. $Q_p$ for a rocky planet is probably $ \sim 10^2$. Such a value would yield much larger tidal heating than illustrated here. Similarly, any transiting planet whose observed radius seems to require the presence of a large, rocky core may be a candidate for severe tidal heating \cite{Jacksonetal08b}. This heating might affect the planet's radius and physical state in ways that have not yet been modeled.\\ 
\indent XO-2 b does not have an inflated radius, which suggests that $e < 0.01$. Otherwise, according to Figs. \ref{Fig5}, there might have been enough recent tidal heating to have pumped up the radius. For XO-4 b, the range of eccentricities we consider does not allow heating to exceed $10^{18}$ W, but the radius has been reported to be inflated \cite{McCulloughetal08}. To explain the large radius by tidal heating would require the eccentricity of XO-4 b to exceed 0.03 or its $Q_p$ value to be smaller than we have assumed. Further study of this planet might include consideration of whether the transit data admit $e > 0.03$. 

\subsection{Planets with reported non-zero eccentricities}
\begin{figure}[h]
\begin{center}
\includegraphics[width=3.4in]{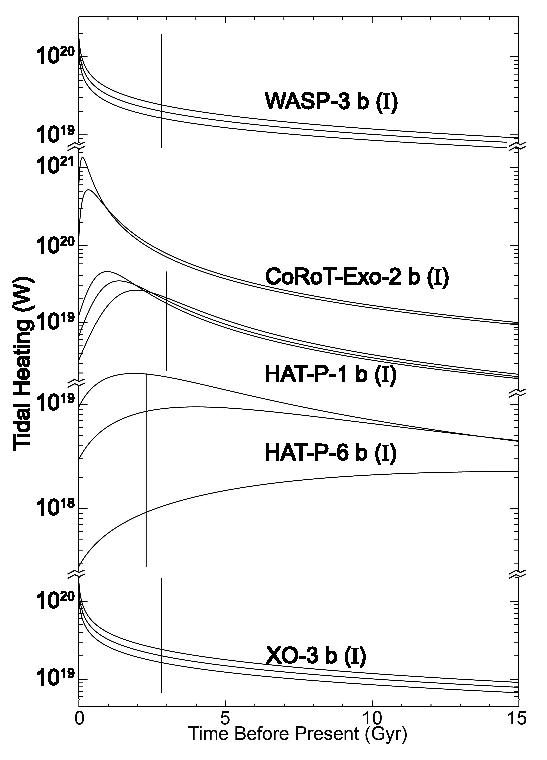} 
\caption{Tidal heating of WASP-3 b, CoRoT-Exo-2 b, HAT-P-1 b, HAT-P-6 b, and XO-3 b. Orbital and physical parameters are taken from \cite{Pollaccoetal08}, \cite{Alonsoetal08}, \cite{Bakosetal07a}, \cite{Noyesetal08}, and \cite{Winnetal08}, respectively. }
\label{Fig6}
\end{center}
\end{figure}

\begin{figure}[h]
\begin{center}
\includegraphics[width=3.4in]{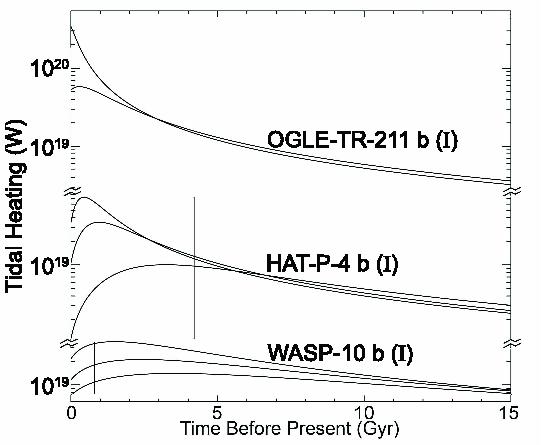} 
\caption{Tidal heating of OGLE-TR-211 b, HAT-P-4 b and WASP-10 b. Orbital and physical parameters taken from \cite{Udalskietal08}, \cite{Kovacsetal07}, and \cite{Christianetal08}, respectively. }
\label{Fig7}
\end{center}
\end{figure}

\indent In Figs. \ref{Fig6} and \ref{Fig7}, we illustrate the tidal heating histories for planets for which some non-zero eccentricity values have been reported. We consider the range of tidal heating allowed by the observational uncertainties of $a$ and $e$. For each of these planets, the heating curves are shown for nominal current $a$ and $e$ values and for sets of $a$ and $e$ (within the range of observational uncertainty) that give the maximum and minimum current tidal heating rates. For CoRoT-Exo-2 b and OGLE-TR-211 b, the uncertainty extends to zero eccentricity, so the corresponding minimum heating rates are zero. Note that, in many other cases as well, the sources of orbital elements suggest that their observations may be consistent with a circular orbit, in which case past and present tidal heating would be small. Heating histories will become more reliable as future observational work yields better determination of orbital eccentricities.

\indent All of the planets in Figs. \ref{Fig6} and \ref{Fig7} are reported to have inflated radii. In each case, the tidal heating exceeds $10^{19}$ W for some allowed value of the eccentricity. These results are consistent with our suggestion that tidal heating, either recent or current, may be responsible for the inflated radii seen in many planetary transits.

\section{Transit Probabilities}
\indent The geometric probability for a planet to transit its host star increases the closer a planet is to the star. Consequently, the transit probability, $P$, is related to the planet's orbital semi-major axis and eccentricity, as given by \cite{Barnes07}:
\begin{equation*}
P = \frac{R_* + R_p}{a (1-e^2)}
\end{equation*}
where $R_*$ is the stellar radius. This expression shows that we are more likely to observe planets transit when they have small $a$ and/or large $e$. Because tidal evolution changes $a$ and $e$, it also changes the transit probabilities of tidally evolving planets. The change in transit probability depends on a planet's specific trajectory through $e$-$a$ space and thus depends sensitively on the physical and orbital parameters of the system. \\
\indent As a result, tidal evolution may introduce certain biases in transit statistics that will become more apparent as more transiting planets are discovered, as we describe below. 

\subsection{Evolution of Transit Probabilities}
\begin{figure}[h]
\begin{center}
\includegraphics[width=3.4in]{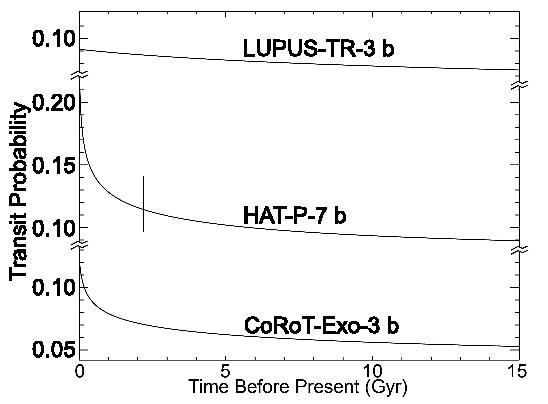} 
\caption{Time evolution of transit probabilities for LUPUS-TR-3 b, HAT-P-7 b, and CoRoT-Exo-3 b. Orbital and physical parameters taken from \cite{Weldrakeetal08}, \cite{Paletal08}, and the exoplanets catalog located at exoplanets.eu, respectively.}
\label{Fig8}
\end{center}
\end{figure}

\indent Applying our tidal evolution calculations (e.g. Fig. \ref{Fig1}) to Eqn. (1) above yields the evolution of transit probabilities. For example, Fig. \ref{Fig8} illustrates the history of transit probabilities for three transiting planets. For LUPUS-TR-3 b, the transit probability is nearly constant over time, while the others show dramatic increases as the present time is approached. The difference arises from the relative contribution of tides raised on the stars. For close-enough-in planets, because a tide raised on a star exerts a negative torque on the planet's orbit, the orbital angular momentum $L$ drops with time. The denominator in Eqn. (1) is proportional to $L^2$, so the transit probability increases with time. The magnitude of the effect is greater for planets with larger masses (because they raise larger tides on the star) and for stars with larger radii. LUPUS-TR-3 b has a much smaller mass relative to its star than CoRoT-Exo-3 b, and its star has a much smaller radius than HAT-P-3. As a result, the tide raised on LUPUS-TR-3 exerts a smaller torque, and $L$ is nearly constant, which explains the constant probability in Fig. \ref{Fig8}.

\subsection{Masses and Ages of Transiting Planets}
\begin{figure}[b]
\begin{center}
\includegraphics[width=3.4in]{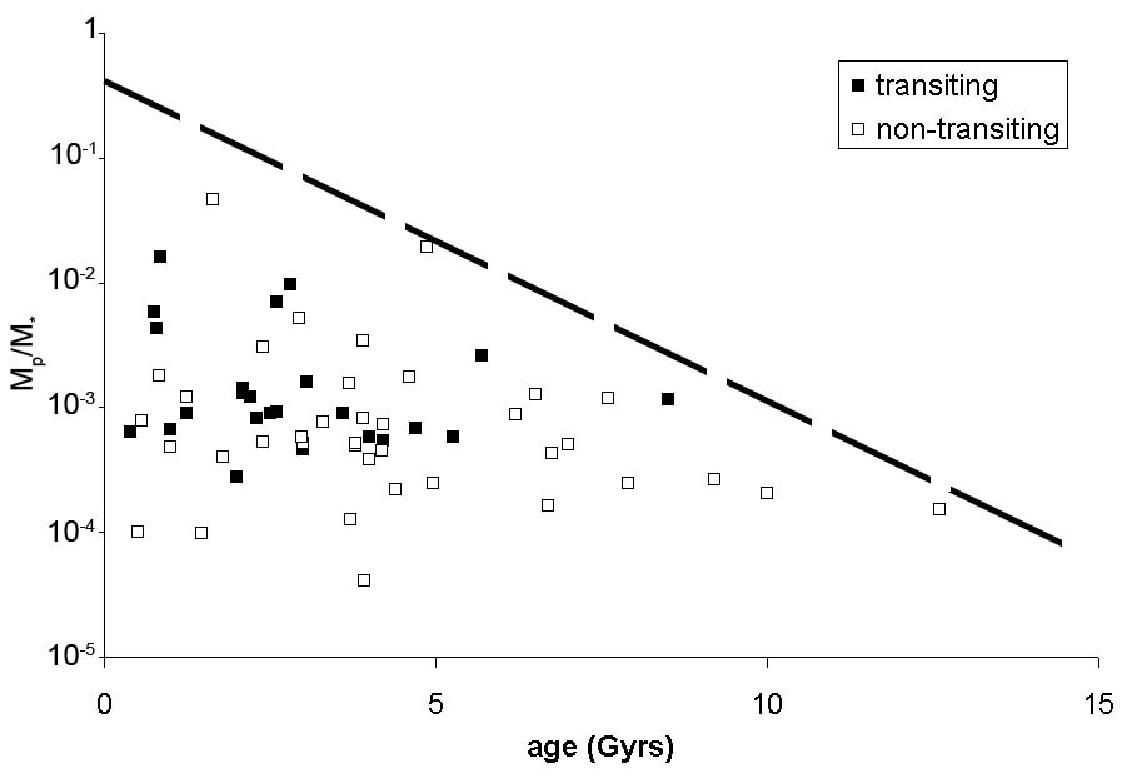} 
\caption{Distribution of planetary mass ratios and stellar ages for planets with a $<$ 0.2 AU. Black squares represent transiting planets, open squares non-transiting planets. The dashed line is drawn by hand but appears to define a region that contains no extra-solar planets. Data are taken from the sources listed in previous captions and from \cite{Butleretal06}, \cite{Bonfilsetal07}, \cite{GnS98}, \cite{Bakosetal07b}, \cite{FnV05}, \cite{Geetal06}, \cite{DaSilvaetal06}, \cite{Nutzmanetal08}, \cite{Johnsonetal06}, \cite{Knutsonetal07}, \cite{Pepeetal04}, \cite{Gillonetal07}, \cite{Holmanetal06}, and \cite{Burkeetal08}.}
\label{Fig9}
\end{center}
\end{figure}

\indent This result has important implications for the physical properties expected for transiting planets. Because the transit probabilities of planets that raise large tides on their stars increase with time, we might expect that transiting planets will tend to be more massive than non-transiting planets, and we might see them more often orbiting stars with larger radii.\\
\indent Fig. \ref{Fig8} seems to suggest that transiting planets will be found preferentially when they are older, after tidal evolution has enhanced the transit probabilities. To test whether this trend is evident in the observed population, we compare all transiting planets to all non-transiting planets for which there is some estimate of the stellar age. Fig. \ref{Fig9} shows the ratio of the planet's mass ($M_p$) to its host star's mass ($M_*$) vs. the best estimate of the stellar age. 

\indent Only one transiting planet is older than 6 Gyr (XO-5 b), whereas there are substantial fraction of non-transiting planets from 6 Gyr old to $>$ 10 Gyr. This result runs contrary to the idea that we would preferentially observe older transiting planets.\\ 
\indent This result can be explained by the fact that, while tidally-evolved, close-in planets have high probabilities to be transiting, this condition may be short-lived as the tides raised on the star cause a planet's semi-major axis to drop faster the closer a planet is to the star. Planets that are close enough to have large transit probabilities have limited lifetimes (\cite{Jacksonetal08d}). Very quickly, tides drag a potentially transiting planet into the Roche zone of the star and tear the planet apart. \\
\indent The dearth of planets above the dashed line in Fig. \ref{Fig9} also corroborates this scenario. The effects of the tide raised on the star are greater for planets with larger mass ratios, so these planets have shorter lifetimes. For example, a typical planet with a mass ratio $\geq$ $10^{-3}$ may not last more than 10 Gyr because Fig. \ref{Fig9} shows a distinct lack of such planets. \\
\indent Fig. \ref{Fig9} also shows several young transiting planets with mass ratios clustered around $10^{-2}$. For planets with such a large mass ratio, the effects of the tide on the star will preferentially enhance the transit probabilities, as discussed in the previous section. The larger fraction of transiting planets relative to non-transiting planets in this region of the figure is also evidence of the potential bias that tidal evolution introduces into transit observations.\\
\indent Fig. \ref{Fig9} shows a lower cut-off in the mass-ratios of transiting planets at about $3 \times 10^{-4}$, while the mass ratios of non-transiting planets extend down to much lower values. However, for the non-transiting planets, their mass values come from radial-velocity observations, and are thus minimum masses. The transiting population probably gives a more accurate representation of the actual distribution of planetary masses. The difference between the two populations may give a sense of the inclination of the orbital planets of the non-transiting planets. The orbits of planets with $M_p/M_* < 10^{-4}$ are likely significantly inclined relative to our line of sight. Moreover, these results suggest the processes of planet formation tend to form hot Jupiters with a planet-to-star mass ratio $10^{-3}$. We encourage theorists to develop models that explain this mass ratio.

\section{Conclusions}
\indent Tidal evolution of the orbits of close-in planets involves coupling between semi-major axis and eccentricity. The effects of tides on both the host star and planet can be important and should be considered. Many of the recently discovered transiting planets have likely undergone significant tidal evolution, and in most cases, both changes in semi-major axis and the effects of the tide on the star have made important contributions to the evolution of orbital eccentricities. Observers should not discount the full range of possible current eccentricities based on overly simplified theories.\\ 
\indent Significant tidal heating accompanied the orbital evolution for most recently discovered transiting planets. In many cases, recent tidal heating rates have been large. These examples support our earlier suggestion that tidal heating may explain the otherwise surprising large radii of close-in planets. In cases for which the planet's radius seems unaffected by heating, we may be able to place upper limits on the eccentricity. In cases for which heating may be required to account for a planet's anomalously large radius, we may be able to place lower limits on the eccentricity. \\
\indent Tides raised on the star by a planet also increase the transit probability over time. The increase in probability depends on the physical and orbital parameters of the system, so tidal evolution likely introduces biases into transit observations. We may be more likely to detect the transits of planets with large masses relative to their stars and planets whose host stars have large radii. Current transit statistics seem to bear out these predictions. Transiting planets tend to have greater masses (relative to their stars) than the minimum mass values for radial-velocity planets, a result indicative of the distribution of orbital inclinations relative to our line of sight. As planets age and move closer to their star, their transit probability spikes, but only for a short time before the planet is destroyed, because close-in planets experience the strongest tidal effects. This process may explain why transiting planets tend to be younger than non-transiting planets. Tidal evolution is a key process in developing the orbital, physical, and statistical properties of planetary systems.

\end{document}